\documentclass[twocolumn,aps,floats,floatfix,prl,superscriptaddress,tightenlines,
showpacs]{revtex4}

\usepackage{bm}
\usepackage{amsmath}
\usepackage{hyperref}

\newcommand{\beq}{\begin{equation}}
\newcommand{\eeq}{\end{equation}}
\newcommand{\bea}{\begin{eqnarray}}
\newcommand{\eea}{\end{eqnarray}}
\newcommand{\OMIT}[1]{}
\newcommand{\bF}{{\bf F}}
\newcommand{\bx}{{\bf x}}
\newcommand{\bv}{{\bf v}}

\begin{document}

\title{Comment on ``Finite Size Corrections to the Radiation Reaction Force in Classical Electrodynamics''}

\author{P.~Forg\'acs$^{1,2}$, T.~Herpay$^{1}$, P.~Kov\'acs}
\affiliation{Wigner Research Centre for Physics, RMKI, H-1525 Budapest 114, P.O.Box 49, Hungary,\\
$^2$LMPT, CNRS-UMR 6083, Universit\'e de Tours, Parc de Grandmont,
37200 Tours, France}

\pacs{41.60.-m, 03.50.De}

\maketitle

In Ref.\ \cite{galley} effective field theory methods have been employed 
to compute the equations of motion of a spherically symmetric charged shell of radius $R$, taking into account the radiation reaction force exerted by the shell's own electromagnetic field up to ${\cal O}(R^2)$. 
The authors of Ref.\ \cite{galley} have stated that the known result for
the self-force of the shell as can be found from Eq.\ (16.28) of the textbook of Jackson \cite{jackson} (see also Chap.\ 4 in the review of Pearle \cite{pearle})
\bea\label{classical}
\bF_{\rm self} &=&  \frac{e^2}{3R^2}(\bv(t-2R)-\bv(t)) \nonumber\\
&=& \frac23 e^2 \left(- \frac1R \ddot{\bx} +  \dddot{\bx} -\frac23 R \bx^{(4)}+ \cdots \right)
\eea
is incorrect, in that the term linear in $R$ should be absent.
We claim that this conclusion of Ref.\ \cite{galley} is incorrect,
and that the textbook result, Eq.\ \eqref{classical} does hold.

First of all we note that in his monumental work \cite{nodvik} Nodvik has derived the equations of motion of an extended, rigid, spherically symmetric charge distribution of radius $R$ taking into account its radiation reaction force (or self-force) up to ${\cal O}(R^2)$ in a manifestly relativistically covariant formalism.
A first inspection of Ref.\ \cite{nodvik} shows that there are in fact
terms linear in $R$ in the equations of motion.
In the special case of a charged shell, the four-vector of the self-force contains a term proportional to $R$, given as
\beq\label{f1shell}
F_{1,{\rm sh}}^\mu = \frac23e^2 R \left[2 \dot x^{\mu}(\ddot
  x_\nu \dddot x^\nu ) +
\ddot x^{\mu}(\ddot x_\nu \ddot x^\nu ) - \frac23 x^{(4)\nu}\right]\,,
\eeq
and the nonrelativistic limit of \eqref{f1shell} yields immediately the textbook result Eq.\ \eqref{classical}.

Independently of Ref.\ \cite{nodvik} we have rederived the
radiation reaction force of an extended, rigid, spherically symmetric
charge distribution and we are in full agreement with Nodvik's results,
and in particular with Eq.\ \eqref{f1shell}.

Pearle has derived an integral representation of the self-force of an extended, rigid, spherically symmetric charge distribution, Eq.\ (9.26) of Ref.\ \cite{pearle}.
We have also performed a direct computation
starting from Eq.\ (9.26) of Ref.\ \cite{pearle} up to ${\cal O}(R^2)$ to check the correctness of our results. We have reproduced Eq.\ \eqref{f1shell} and at the same time we could also confirm the correctness of the ${\cal O}(R^2)$ term computed in Ref.\ \cite{galley}.
Since the results of Ref.\ \cite{pearle} all fit in the rather general framework of Nodvik, it is only to be expected that the results should agree.

Galley et al.\ argue that the most general world-line action of an extended shell,
Eq.\ (17) of Ref.\ \cite{galley} cannot contain ${\cal O}(R)$ terms, since they are excluded by Poincar\'e and gauge symmetries together with reparametrization invariance of the world-line action.
This argument is correct for terms containing the gauge field, $A_{\mu}$,
however the effective action of an extended charged object does contain a term {\sl not explicitly depending on} $A_{\mu}$ and being of ${\cal O}(R)$.
The appearance of such a term is due to the relativistic kinematics of an
extended rigid body as it is well exposed in Ref.\ \cite{nodvik},
where also the effective action is given. From Eq.\ (7.73)
of Ref.\ \cite{nodvik} one can obtain the ${\cal O}(R)$ term for a shell in the action, given as
\beq\label{R}
S^{(1)}= -\frac{2}{9}R e^2 \int \, d\tau a^2(\tau)\,.
\eeq
As one can see $S^{(1)}$ does not depend explicitly on $A_\mu$ and it 
is consistent with all symmetry requirements.

In fact one can learn from Refs.\ \cite{dirac,bhabha} that the
radiated four momentum receives contributions only from the
difference of the advanced and the retarded potentials, which
is an even function of $R$.
Therefore it is not surprising that Galley et al.\ could match their result
(Eq.~(22) of Ref.\ \cite{galley}) based on an incomplete effective action for the radiated power to the result of Ref.\ \cite{marengo}.
However, the total four-momentum of the self-field generated by the shell,
contributing to the self-force in the equations of motion
contains a ``bound'' part depending of both even an odd powers of the
shell's radius \cite{teitelboim}.

This work has been supported by OTKA grants NI68228 and K101709.

\end{document}